\documentclass[aps,amsfonts,prl,twocolumn,showpacs]{revtex4-1}

\makeatletter
\AtBeginDocument{\let\LS@rot\@undefined}
\makeatother

\usepackage{subfigure}
\usepackage{textcomp}
\usepackage{graphicx}
\usepackage{amssymb}
\usepackage{amsmath}
\usepackage{bm}
\usepackage{amsmath, amsthm, amssymb}
\usepackage{dsfont}
\usepackage[colorlinks=true,linkcolor=blue,urlcolor=blue,citecolor=blue]{hyperref}
\usepackage{multirow}
\usepackage{url}

\def\beq{\begin{equation}}
\def\eeq{\end{equation}}
\def\bea{\begin{eqnarray}}
\def\eea{\end{eqnarray}}

\usepackage[usenames, dvipsnames]{color}

\usepackage{pdfpages} 
\makeatletter
\AtBeginDocument{\let\LS@rot\@undefined}
\makeatother

\begin{document}

\title{Hydrodynamics of nonintegrable systems from a relaxation-time approximation}

\author{Javier Lopez-Piqueres$^{1}$, Brayden Ware$^{1}$, Sarang Gopalakrishnan$^{2}$, and Romain Vasseur$^{1}$}
\affiliation{
$^1$Department of Physics, University of Massachusetts, Amherst, Massachusetts 01003, USA \\
$^2$Department of Physics and Astronomy, CUNY College of Staten Island, Staten Island, NY 10314;  Physics Program and Initiative for the Theoretical Sciences, The Graduate Center, CUNY, New York, NY 10016, USA}

\begin{abstract}

%
We develop a general kinetic theory framework to describe the hydrodynamics of strongly interacting, nonequilibrium quantum systems in which integrability is weakly broken, leaving a few residual conserved quantities. 
This framework is based on a generalized relaxation-time approximation; it gives a simple, but surprisingly accurate, prescription for computing nonequilibrium transport even in strongly interacting systems. 
We validate the predictions of this approximation against matrix product operator calculations on chaotic quantum spin chains, finding surprisingly good agreement.
We show that despite its simplicity, our framework can capture phenomena distinctive to strongly interacting systems, such as widely separated charge and energy diffusion constants.

\end{abstract}

\maketitle

Hydrodynamics has experienced a revival in the past decade, as an effective theory of strongly interacting quantum matter far from equilibrium~\cite{2015arXiv151103646C,
2018arXiv180509331G,2017arXiv170208894L,PhysRevB.97.035127,kvh, rpv,PhysRevLett.122.091602,PhysRevX.9.041017,2020arXiv200309429G,2020arXiv200405177R}. A major factor in this revival has been the advent of new experimental platforms, from quark-gluon plasmas~\cite{2017arXiv171205815R} to strongly interacting ultracold gases~\cite{sommer2011universal, cao2011universal} and pristine solid-state systems that feature strong interactions and long mean free times~\cite{Bandurin1055,Crossno1058,Moll1061,Sulpizio:2019aa}.  Hydrodynamics is particularly rich for low-dimensional fluids, featuring transport anomalies such as long-time tails~\cite{PhysRevLett.18.988,Ernst:1984aa,PhysRevB.73.035113,PhysRevA.89.053608, spohn1991large, delacretaz2020breakdown}; in \emph{one} dimension, hydrodynamics is further enriched by the proximity of many realistic systems to integrability. In the integrable limit, conventional hydrodynamics breaks down, and a new framework, called ``generalized hydrodynamics'' (GHD), has been developed~\cite{Doyon,  Fagotti, SciPostPhys.2.2.014, PhysRevLett.119.020602,  BBH0, BBH,PhysRevLett.119.020602, piroli2017, GHDII, doyon2017dynamics, solitongases,PhysRevLett.119.195301, PhysRevB.96.081118,PhysRevB.97.081111,alba2017entanglement, PhysRevLett.120.176801, dbd1, ghkv, dbd2, PhysRevB.100.035108,PhysRevLett.122.090601, 2019arXiv190601654B, pozsgay2019current, 2019arXiv190807320B, 2020arXiv200407113Y, pozsgay2020algebraic,10.21468/SciPostPhys.8.3.041,bertini2020finitetemperature}. 
GHD incorporates the distinctive features of integrable dynamics: namely, the presence of infinitely many conservation laws and of stable ballistically propagating quasiparticles. This framework has led to quantitative explanations of many phenomena, including Drude weights~\cite{PhysRevLett.119.020602,BBH,
PhysRevB.96.081118,GHDII}, diffusion constants~\cite{dbd1,ghkv, dbd2,gv_superdiffusion,2019arXiv191101995M,2019arXiv191201551D}
 and the presence of anomalous transport in strongly interacting spin chains~\cite{1742-5468-2009-02-P02035,lzp,sanchez2018anomalous, idmp, gv_superdiffusion,gvw,PhysRevLett.123.186601,PhysRevLett.122.210602, manas2019, 2019arXiv190905263A,2019arXiv191008266B,2020arXiv200305957K,nardis2020superdiffusion,2020arXiv200505984F}.

Realistic systems, however, are only \emph{approximately} integrable. On short timescales they obey GHD, but on the longest timescales they cross over to conventional hydrodynamics. A general theory of this crossover has remained elusive, despite recent progress~\cite{landau1981course,PhysRevLett.111.197203, huangkarrasch, PhysRevB.89.165104, PhysRevLett.96.067202, PhysRevA.89.053608, Langen_2016, PhysRevX.5.041043, PhysRevLett.115.180601, PhysRevB.94.245117,2017arXiv171100873C, PhysRevE.81.036206,  PhysRevLett.120.070603, PhysRevB.94.214304, PhysRevLett.119.010601, sanchez2018anomalous, PhysRevLett.123.240603, Biella:2019aa, PhysRevLett.120.164101,  fgv,2020arXiv200411030D, de2020universality,bastianello_dephasing, pandey2020adiabatic,brenes2020eigenstate,brenes2020ballistic}. 
In principle one can write a collisional Boltzmann equation for weak integrability breaking~\cite{fgv,2020arXiv200411030D}. However, in general the collision integral is intractable, as it depends on \emph{all} the matrix elements of the integrability-breaking perturbation. In special cases, such as long-range interactions, slowly fluctuating noise, or weakly interacting systems, the integrability-breaking perturbation can itself be expressed in terms of GHD data~\cite{fgv,2020arXiv200411030D}. More generally, however, integrability-breaking perturbations lie outside GHD: for example, umklapp scattering involves large momentum transfer, and thus cannot be captured by a long-wavelength theory such as GHD. In the absence of the GHD framework, evaluating the collision integral is an intractable task. 

This work addresses the question of integrability breaking from a fundamentally different perspective. Instead of microscopically deriving the collision integral, we adopt a simple but general approximation, which we call the ``generalized relaxation time approximation'' (GRTA), by analogy with the conventional relaxation time approximation (RTA) for weakly interacting electrons~\cite{brennan1999physics}. The GRTA assumes that there is a single dominant relaxation time that controls the onset of chaos. 
This assumption allows us to efficiently simulate dynamics away from the integrable limit. 
Although our approach resembles the conventional RTA in positing a unique relaxation time, its implementation and physical consequences are completely different. The RTA deals with nearly free particles, so their scattering kinematics is simple. By contrast, in an interacting integrable system, the momentum carried by each quasiparticle is a nonlinear functional of the full quasiparticle distribution function. Thus, when one describes a scattering process in an integrable system, not only the matrix elements but also the delta functions conserving momentum and energy are nontrivial to evaluate. 

Instead, we implement the GRTA as follows. In GHD, one regards a system as locally being in a generalized Gibbs ensemble (GGE)~\cite{Rigol:2008kq,Langen207,1742-5468-2016-6-064007}, with chemical potentials for each conservation law~\cite{1742-5468-2016-6-064007}. The key step in our approach is to replace the local GGE with a \emph{local} thermal Gibbs state, subject to the residual conservation laws, at some finite rate $1/\tau$ (where $\tau$ is the generalized relaxation time). The main assumption is that there is a unique \emph{local} relaxation rate for the quasiparticle distribution function. This is justified under certain assumptions, and (as we discuss below) fails sometimes; however, we find that it is remarkably accurate at reproducing numerical time evolution, even when the integrability-breaking perturbations are not especially small. For initial states far from equilibrium, the GRTA (unlike the RTA) gives rise to nontrivial relaxation dynamics, as the local equilibrium state is a nontrivial functional of the local quasiparticle distribution. Moreover, contrary to the simplest implementation of the RTA, GRTA preserves conservation laws and is suitable to study hydrodynamics. Thus, we argue the GRTA captures the ``generic'' crossover from generalized to conventional hydrodynamics. 

{\bf Boltzmann equation. }GHD describes the dynamics of integrable systems in terms of their quasiparticles. We characterize quasiparticles with a given quantum number (``rapidity'') $\lambda$ by their density $\rho^{~}_\lambda(x,t)$. Note that $\lambda$ is a shorthand for both continuous and discrete labels. The distribution of quasiparticles $\rho^{~}_\lambda(x,t)$ is in one-to-one correspondence with a local equilibrium macrostate~\cite{1742-5468-2016-6-063101}.
In an integrable system with conserved charges $\lbrace \hat{Q}_n \rbrace$,  local equilibrium can be equivalently characterized by a generalized Gibbs ensemble (GGE) density matrix $\hat{\rho}_{\rm GGE} = Z^{-1} {\rm e}^{- \sum_n \beta_n \hat{Q}_n}$. In integrable systems, quasiparticles scatter elastically with phase shifts leading to Wigner time delays~\cite{solitongases,BBH}:
 the effective velocity  $v^{\rm eff}_\lambda [\rho] $ of a quasiparticle with rapidity $\lambda$ depends on the density of all the other quasiparticles~\cite{PhysRevLett.113.187203,Doyon,  Fagotti,2019arXiv190807320B}.
Transport properties can be inferred from the fact that quasiparticles carry some charge $h_i(\lambda)$, where $i$ labels the conserved charges of the integrable system. The density of charge $i$ reads $q_i(x,t) = \int d\lambda  h_i(\lambda)  \rho_\lambda(x, t)$,  with the associated Euler current $j_i(x,t) = \int d\lambda  h_i(\lambda)  \rho_\lambda(x, t) v^{\rm eff}_\lambda [\rho] +\dots$, where ``$\dots$'' represents higher order (diffusive) corrections~\cite{dbd1, ghkv, dbd2,2019arXiv191101995M} 
that will be negligible for our purposes. The conservation laws $\partial_t q_i + \partial_x j_i=0$ form the basis of GHD~\cite{Doyon,Fagotti}. 

We now imagine perturbing such an integrable system with Hamiltonian $\hat{H}_0$ by a small, nonintegrable perturbation $ \hat{V}$ of order $g$ that destroys all but a few conservation laws. We assume that the expressions for charges and currents are unchanged -- neglecting ${\cal O}(g)$ corrections to these quantities, and force terms that are treated elsewhere~\cite{bastianello_inhomogeneous}. The leading effect of the non-integrable perturbation is to thermalize quasiparticle distributions at long times $t \gg  {\cal O}(g^{-2})$. Integrability breaking endows the GHD equation with a collision integral
\begin{equation}
\partial_t \rho_\lambda + \partial_x \left( v^{\rm eff}_\lambda [\rho]  \rho_\lambda \right) = \mathcal{I}_{\lambda} [\rho].~~~ \label{eqBoltzmannGHD}
\end{equation}
that mixes quasiparticle sectors. 
This collision integral $\mathcal{I}_{\lambda}$ can in principle be derived perturbatively using Fermi's Golden Rule (FGR), and is ${\cal O}(g^{2})$~\cite{fgv,2020arXiv200411030D,bastianello_dephasing}. It involves the matrix elements (form factors) of the integrability breaking perturbations, which can be expressed in terms of hydrodynamical data only for noninteracting systems, and for perturbations involving low momentum transfer such as slowly varying noisy potentials or long-range interactions~\cite{fgv}. Eq.~\eqref{eqBoltzmannGHD} was analyzed within linear response in Ref.~\cite{fgv}, and was shown to lead to diffusive hydrodynamics in general.

{\bf Generalized relaxation-time approximation. } For most physical integrability-breaking perturbations, the matrix elements of the perturbation cannot be expressed in terms of hydrodynamic data.  In the few cases where the collision integrals can be written down explicitly, they are impractical to implement numerically, even for simple physical processes like particle loss in a Bose gas~\cite{10.21468/SciPostPhys.9.4.044}. For context, we remark that even for weakly-interacting fermions, collision integrals are often approximated by using the relaxation-time approximation (RTA), which suffices to capture most of the relaxation physics and to describe experiments. Here, we introduce a generalized relaxation-time approximation (GRTA), which amounts to choosing a simple form for the collision integral:
\begin{equation}
\partial_t \rho_\lambda + \partial_x \left( v^{\rm eff}_\lambda [\rho]  \rho_\lambda \right) = - (\rho_\lambda - \rho^{\rm Gibbs}_\lambda[\rho]  )/\tau.~~~ \label{eqGRTA}
\end{equation}
This right-hand side enforces local thermalization on a typical relaxation timescale $\tau$ as follows: $\rho^{\rm Gibbs}_\lambda[\rho]$ is a nonlinear functional of the state $\rho_\lambda$, defined as the distribution of quasiparticles of a Gibbs state with the same value of the conserved quantities $q_\alpha$ ($\alpha=1,\dots,N$ corresponding to the charges preserved by the integrability breaking perturbation) as the state $\rho_\lambda$. For example, consider a Bose gas where the integrability breaking perturbation preserves energy $E$, particle number $N$ and momentum $P$. Then the distribution $\rho^{\rm Gibbs}_\lambda[\rho] $ corresponds to the (boosted) Gibbs ensemble density matrix $\hat{\rho}_{\rm Gibbs}  = \frac{1}{Z} {\rm e}^{- \beta (\hat{H} -\mu \hat{N} - \nu \hat{P})}$ where $\beta, \mu$ and $\nu$ are chosen so that the average particle number, energy and momentum are the same as in the state $\rho_\lambda$. By definition, we have $\int d\lambda (\rho_\lambda - \rho^{\rm Gibbs}_\lambda)h_\alpha(\lambda) = 0$, ensuring the conservation of the charges $\hat{Q}_\alpha$. 

Physically, the GRTA assumes that local relaxation is controlled by a single relaxation rate. Of course, realistic FGR collision integrals have a lot more structure, involving a hierarchy of relaxation rates. However, we expect this approximation to capture the key physics of integrability breaking. One can formalize this intuition as follows. The relaxation of charges in the presence of weak integrability-breaking is captured by the equation $\partial_t Q_i = - \sum_j \Gamma_{ij} Q_j$, where ${\bf \Gamma}$ is a matrix that is itself a functional of the equilibrium state~\cite{fgv, 2020arXiv200411030D}. 
%
%
The spectrum of  the matrix ${\bf \Gamma}$ contains zero modes corresponding to the residual conserved charges, as well as other eigenmodes that capture the characteristic decay rates. 
If there is a gap between the zero modes and the decaying modes, one can identify this gap with $1/\tau$, and replace the matrix ${\bf \Gamma}$ with a projector onto modes that decay at rate $\sim 1/\tau$, which is justified at long enough times where ${\rm e}^{- t /\tau}$ will dominate exponentials decaying with faster rates. 
The GRTA corresponds to replacing ${\bf \Gamma}^{-1} \approx \tau$ for all decaying charges, which approximately coincides with the projection approach, provided that all residual conserved currents have approximately similar overlaps with the slowest-decaying modes of $\mathbf{\Gamma}$. (This construction indicates that the GRTA will fail whenever there are arbitrarily slowly relaxing modes, as we expect on physical grounds, and also when the currents of residual charges have very different overlaps with the slowest-relaxing modes of ${\bf \Gamma}$.) 

We evaluate the right-hand side of eq.~\eqref{eqGRTA} as follows. We compute the (density of) conserved charges $q_\alpha$ (say particle number, momentum and energy) in the state $\rho_\lambda(x,t)$, and invert the equation of states of the model -- known from the equilibrium thermodynamic Bethe ansatz (TBA)~\cite{Takahashi} --  
 to find the Lagrange multipliers (in our example, $\beta$, $\mu$ and $\nu$) of the Gibbs state corresponding to those values. Using TBA, we then compute the density of quasiparticles $\rho^{\rm Gibbs}_\lambda[\rho]$ corresponding to those Lagrange multipliers and thus $I_{\lambda}$ \cite{suppmat}. Note that we use the equation of states of the {\em unperturbed} (integrable) model. This is justified perturbatively by the fact that the integrability breaking perturbation smoothly modifies thermodynamic quantities and the equation of states (with small changes if the perturbation is weak), while it dramatically affects the dynamics at long times. 
 We take $\tau$ to be an unknown constant, a single phenomenological parameter to be determined by comparing the solution of eq.~\eqref{eqGRTA} to numerics or experiments.
 
\begin{figure}[t!]
\includegraphics[width=\linewidth,clip]{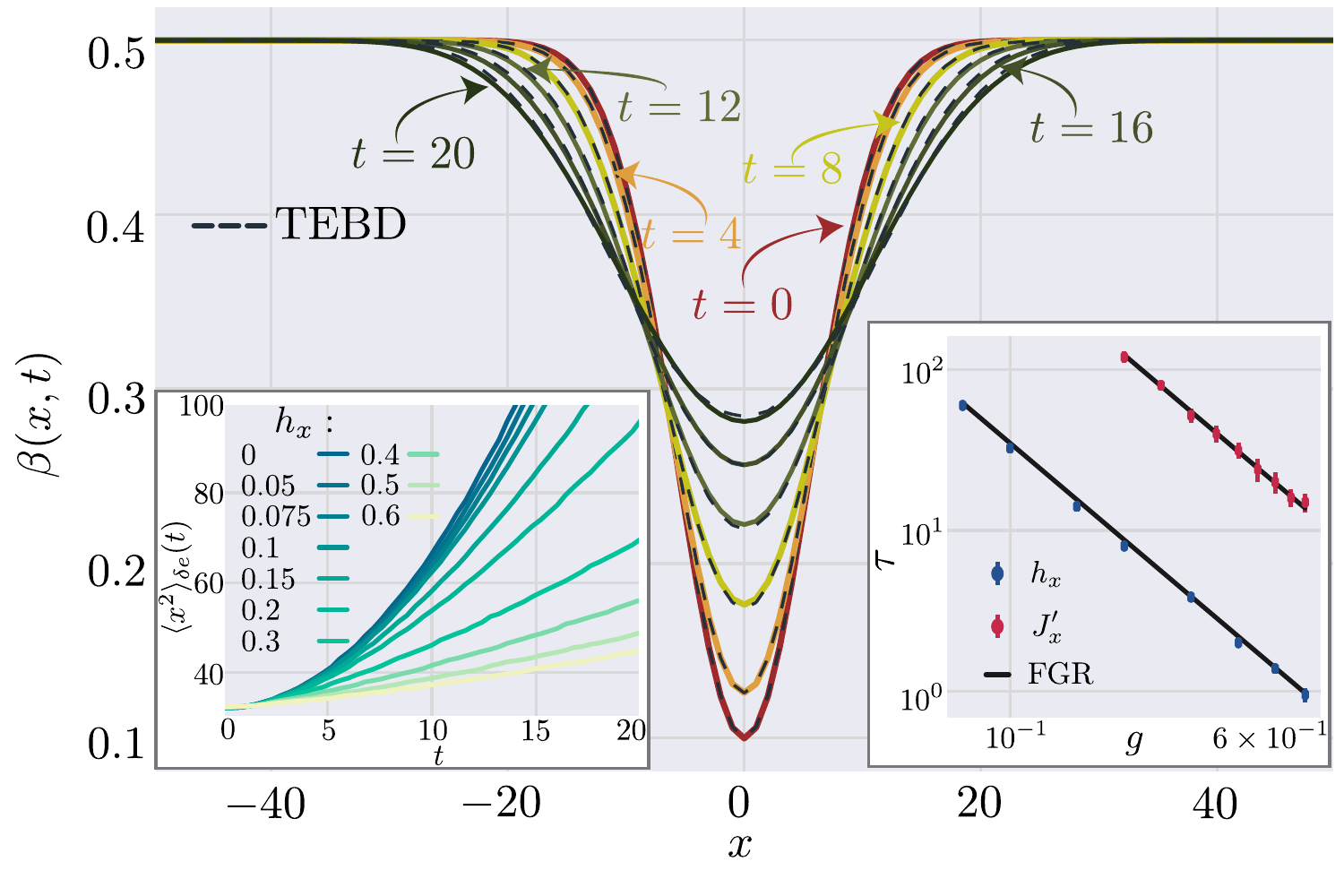}
\caption{{\bf Energy transport in nonintegrable spin chains}: inverse temperature profiles $\beta(x,t)=1/T(x,t)$ in an XXZ spin chain with a staggered transverse field $h_x$ breaking integrability. The TEBD data for $h_x=0.2$ is described very well by eq.~\eqref{eqBoltzmannGHD2} and GRTA with $\tau \simeq 8$. {\it Left inset:} Variances of the energy profiles {\it vs} time from TEBD, for various values of $h_x$, showing a crossover between ballistic and diffusive transport. {\it Right inset:} The fitted values of $\tau$ agree with the FGR scaling~\eqref{eqFRG} for both $g=h_x$ (staggered $x$-fields) and $g=J'_x$ (staggered $xx$-couplings).  }
\label{Fig3}
\end{figure}

{\bf Numerical solution.} To implement this GRTA scheme numerically , we develop a general numerical scheme to solve~\eqref{eqBoltzmannGHD}, which can be used both near and far from equilibrium. Following the numerical methods of Ref.~\cite{BBH0, bastianello_inhomogeneous,10.21468/SciPostPhys.8.3.041}
 in the integrable case, we find it convenient to work with the ``normal modes'' of GHD, which are given by the occupation ratios (Fermi factors) $n_\lambda = \rho_\lambda/\rho^{\rm tot}_\lambda$, where $\rho^{\rm tot}_\lambda = \rho_\lambda +  \rho^h_\lambda$ is the total density of states at rapidity $\lambda$ and $\rho^h_\lambda$ the density of holes. There is a one-to-one correspondence between the density of quasiparticles $\rho_\lambda$ and the occupation ratios $n_\lambda$, provided by the Bethe equations. In terms of $n_\lambda$, the Boltzmann equation~\eqref{eqBoltzmannGHD} takes the {\em advection} form
\begin{equation}
\partial_t n_\lambda + v^{\rm eff}_\lambda [n] \partial_x  n_\lambda  = I_{\lambda} [n],~~~ \label{eqBoltzmannGHD2}
\end{equation}
where $I_{\lambda} $ is simply related to $\mathcal{I}_{\lambda} [\rho]$~\cite{suppmat}. We then solve this equation by finite elements, discretizing space, time, and rapidity. We use a backward first order scheme $n_\lambda(x,t) = n_\lambda(x -  v^{\rm eff}_\lambda [n(x,t)] \Delta t,t-\Delta t) + \Delta t I_{\lambda} [n(x,t)] $, where crucially, the velocity and collision integrals in the right-hand side are evaluated at time $t$ to improve stability. We solve this equation by iteration, and check convergence with respect to the small parameters $\Delta t$, $\Delta x$ and $\Delta \lambda$. 


{\bf Energy transport in spin chains.} The GRTA approach has the advantage of being very general, and can be applied to chaotic spin chains near integrability. To illustrate this, we consider the spin-$\frac{1}{2}$ XXZ spin chain with integrability breaking perturbations
\begin{equation}
\hat{H} = \sum_i (\hat{S}_i^x \hat{S}_{i+1}^x + \hat{S}_i^y \hat{S}_{i+1}^y +\Delta \hat{S}_i^z \hat{S}_{i+1}^z ) + {\hat V},
\label{eqXXZperturbed}
\end{equation}
with anisotropy $\Delta=\frac{1}{2}$, and ${\hat V} = h_x \sum_i (-1)^i  \hat{S}_i^x$ or ${\hat V} =J'_x \sum_i (-1)^i \hat{S}_i^x \hat{S}_{i+1}^x $ . When ${\hat V} = 0$, this model is integrable, and energy transport is purely ballistic as the total energy current is a conserved quantity. As higher-order corrections vanish exactly, energy transport can be captured extremely well by GHD~\cite{BBH0}. The staggered perturbation ${\hat V}$ breaks integrability and the $U(1)$ symmetry of the XXZ model.

We consider energy transport in the Hamiltonian~\eqref{eqXXZperturbed} by preparing a local region with temperature $T=10$ embedded in a uniform equilibrium background with temperature $T=2$.~\footnote{We have chosen energy transport in this model because GHD accurately describes energy transport in the integrable limit even at relatively short times. For other quantities like spin, GHD remains asymptotically valid, but there are larger corrections at short times; these are corrections to GHD, rather than to GRTA.}. 
We simulate the dynamics of this system up to time $t=20$ by evolving the density matrix using time-evolving block decimation (TEBD)~\cite{PhysRevLett.91.147902,vidal,schollwoeck}
 and compare with the GRTA~\eqref{eqGRTA} for various values of $\tau$. We compare the local temperature profiles $T(x,t)$ between the two approaches, using the equilibrium equation of state of Eq.~\eqref{eqXXZperturbed} to convert energy density to temperature. (This accounts for the shift in the equilibrium energy density due to the perturbation ${\hat V}$, which can readily be captured using perturbation theory). We find a best fit for the single parameter $\tau$ by matching the full temperature profiles from the TEBD simulations and the GRTA.

We find that GRTA is able to describe the nonintegrable dynamics of~\eqref{eqXXZperturbed} remarkably well with a single parameter $\tau$ for each ${\hat V}$, for various values of $h_x$ or  $J'_x$ ranging from $0.05$ to $0.6$, corresponding to almost two decades in $\tau$. Moreover, the fitted values of $\tau$ all agree very well with the simple FGR scaling 
\begin{equation} \label{eqFRG}
\tau \simeq C g^{-2},
\end{equation}
with $C \approx 0.32(5)$ for $g=h_x$ (staggered $x$-fields), and  $C \approx 4.95(5)$ for $g=J'_x$ (staggered $xx$-couplings). This is remarkable, as in general we expect that relaxation times should depend on temperature, and the initial state considered has a wide range of temperatures. Allowing for limited dependence of $\tau$ on the state $\rho$ -- such as through the local temperature --- might be necessary to capture strongly nonequilibrium setups with even wider temperature ranges. While the variance of the profiles of the local perturbation in energy grows quadratically (indicating ballistic transport) in the integrable case, it crosses over to linear (diffusive) growth for times $t \gg \tau$.

\begin{figure}[t!]
\includegraphics[width=\linewidth,clip]{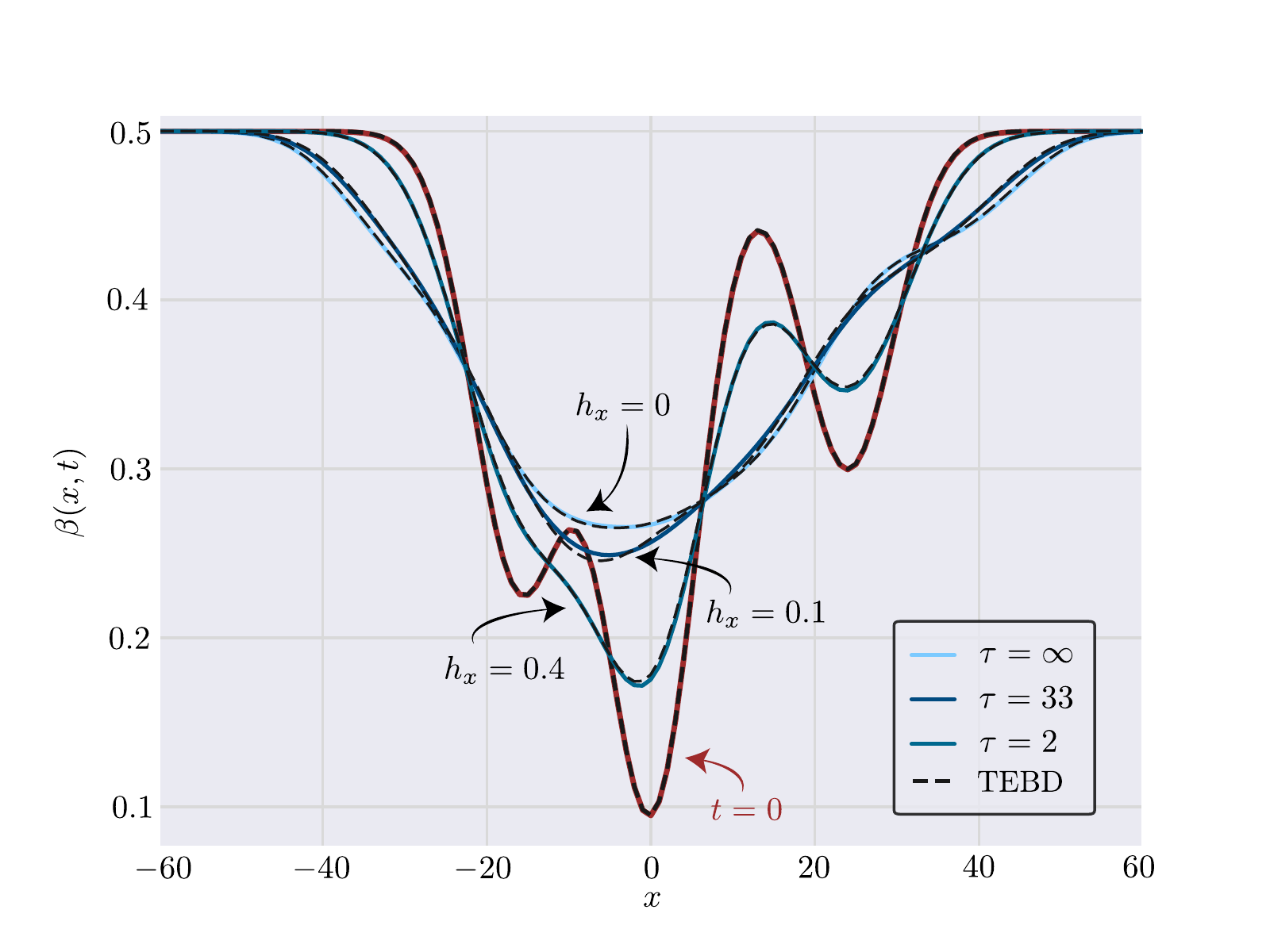}
\caption{{\bf Generic energy transport in chaotic spin chains}: inverse temperature profiles $\beta(x,t)=1/T(x,t)$ at time $t = 20$ in an XXZ spin chain with a staggered transverse field $h_x$ breaking integrability, comparing TEBD and GRTA starting from a non-trivial inhomogeneous initial state. The values of $\tau$ in GRTA for each $h_x$ were determined from Fig.~\ref{Fig3}.   }
\label{Fig3b}
\end{figure}

This scaling implies that the whole time evolution for all values of $g$ we consider can be described quite accurately using a single free parameter $C$. While we obviously expect corrections to this GRTA approach, combined with the expected FGR scaling~\eqref{eqFRG}, it clearly captures most of the physics of integrability breaking. Surprisingly GRTA is able to describe energy transport even for strongly chaotic chains for which the relaxation time $\tau$ is ${\cal O}(1)$. 

To illustrate the predictive power of GRTA, we study energy transport for a more complicated inhomogeneous initial state, for various values of the staggered field $h_x$, comparing GRTA to TEBD (Fig~\ref{Fig3b}). Note that there is no free parameter here, as the values of the relaxation time $\tau(h_x)$ are fixed from the analysis of Gaussian initial states in Fig.~\ref{Fig3}, and follow approximately eq.~\eqref{eqFRG}.  The agreement is remarkable, and illustrates that GRTA captures energy transport in this generic nonintegrable spin chain not only qualitatively, but also to a large extent quantitatively (the error between GRTA and TEBD is at most $2 \%$).   
  
{\bf Hydrodynamics of non-integrable Bose gases.} We also used the GRTA to capture the crossover from generalized to conventional hydrodynamics in one-dimensional Bose gases, described by the Lieb-Liniger model~\cite{suppmat}. We considered integrability-breaking perturbations that either relax or preserve momentum,  and implemented both far from equilibrium free expansions into vacuum of a cloud of atoms which models experiments on ultracold  Bose gases~\cite{kinoshita,
PhysRevB.79.214409,PhysRevB.84.205115,
PhysRevLett.110.205301,
PhysRevB.89.075139,
PhysRevLett.109.110602,
PhysRevA.85.043618,PhysRevB.88.235117,PhysRevB.95.035155,PhysRevB.95.060406,
PhysRevLett.122.090601,tang2018}, and linear response setups where the initial state is a small local perturbation on top of an equilibrium Gibbs state.
For Bose gases, one can consider integrability breaking perturbations that conserve momentum, as well as energy and particle number. We briefly summarize some key findings~\cite{suppmat}. (1)~For perturbations that conserve energy, particle number, and momentum, we recover the hydrodynamics of a conventional fluid, with separate heat and sound peaks. (2)~For perturbations that conserve only energy and particle number, we find well-separated diffusion constants for these two conserved quantities. This is a natural consequence of GHD, since the energy and particle-number Drude weights are different, but illustrates the strongly interacting nature of the dynamics we are able to capture. (In noninteracting systems these quantities would be linked by the Wiedemann-Franz law.) (3)~Finally, although the GRTA assumes a uniform relaxation time $\tau$, starting from nonequilibrium states we find that different charges can approach their equilibrium values at different rates. This is due to the highly nonlinear nature of the GRTA, discussed above.

%

{\bf Discussion.} In this work we have introduced the GRTA as a numerically efficient approximation to study the nonequilibrium dynamics of systems with weak integrability breaking. The GRTA treats integrability-breaking in a rather drastic approximation, where all but the residual conserved charges decay on a single timescale $\tau$. Nevertheless, this approximation works surprisingly well to capture the hydrodynamics of physically relevant integrability-breaking perturbations $ {\hat V}$(such as a staggered transverse field in the XXZ model) at the cost of introducing $\tau(g) = C g^{-2}$ with a single fit parameter $C$. Many natural extensions of this method suggest themselves. For instance, in cases where some charges relax much slower than others, we can treat the dynamics of the fast charges within GRTA (treating the slow modes as conserved) and then relax the slow charges separately. This could be relevant, for example, in ultracold atomic experiments, where integrability breaking due to collisions can be much faster than atom loss or momentum relaxation due to the trap. Another natural extension would be to add noise to the GRTA equations (of strength given by the fluctuation-dissipation theorem). 
Finally, our implementation of the integrable dynamics itself has been restricted to Euler scale hydrodynamics. An important open question is to develop an efficient scheme for numerically solving the GHD equations beyond the Euler scale~\cite{dbd1}; incorporating the GRTA into this scheme would allow us to answer currently open questions about the fate of anomalous diffusion in nonintegrable spin chains~\cite{de2020universality}.

\begin{acknowledgments}

\emph{Acknowledgments}.---The authors thank V. Bulchandani, J. De Nardis and P. Dumitrescu for useful discussions. S.G. and R.V. also thank  A. Friedman for collaborations on related topics. This work was supported by the National Science Foundation under NSF Grant No. DMR-1653271 (S.G.),  the US Department of Energy, Office of Science, Basic Energy Sciences, under Early Career Award No. DE-SC0019168 (R.V. and J.L.),  and the Alfred P. Sloan Foundation through a Sloan Research Fellowship (R.V.). 

\end{acknowledgments}

\bibliography{refs}

\bigskip

\onecolumngrid
\newpage



\section*{Supplementary material}

# Supplemental Material for "Hydrodynamics of nonintegrable systems from a relaxation-time approximation"

Javier Lopez-Piqueres[1], Brayden Ware[1], Sarang Gopalakrishnan[2] and Romain Vasseur[1]

[1]*Department of Physics, University of Massachusetts, Amherst, Massachusetts 01003, USA*
*Department of Physics and Astronomy, CUNY College of Staten Island,*
*Staten Island, NY 10314; Physics Program and Initiative for the Theoretical Sciences,*
*The Graduate Center, CUNY, New York, NY 10016, USA*

## I. BOLTZMANN EQUATION IN NORMAL MODE BASIS

The form of the Boltzmann equation written in eq (1) of the main text is not the most convenient when it comes to solving it numerically. Instead it is simpler to work with the occupation ratios (Fermi factors) or *normal modes* of the theory which are related to the density of quasiparticles via the *local* Bethe equation

$$\rho_\lambda^{\rm tot}(x,t) = \frac{p'_\lambda}{2\pi} + \int d\lambda' T_{\lambda,\lambda'} \rho_{\lambda'}(x,t), \tag{1}$$

with the kernel $T_{\lambda,\lambda'}$ is the scattering kernel of the theory and $p'_\lambda = dp/d\lambda$, with $p_\lambda$ the bare momentum. We can omit the space-time dependence and also write all equations abstractly working directly with operators and vectors acting on rapidity space, see Ref.[1] for a recent review. In this language, the kernel $\hat{T}$ of the integrable model in question is an operator that acts as the convolution when applied to a vector, i.e. $\hat{T}\vec{h}|_\lambda = (T*h)|_\lambda = \int d\lambda' T_{\lambda,\lambda'} h_{\lambda'}$. All the operators we will deal with will be diagonal, e.g. $\hat{n}\vec{h}|_\lambda = n_\lambda h_\lambda$. This way the Bethe equation reads simply $\vec{\rho} = \frac{1}{2\pi} \hat{n} \vec{p'}^{\rm dr}$, where the superscript dr symbolizes *dressing* and for any vector it is simply given by $\vec{h}^{\rm dr} = (\hat{1} - \hat{T}\hat{n})^{-1}\vec{h}$. The GHD equation with a collision integral in this notation thus reads

$$\partial_t \vec{\rho} + \partial_x \left( \hat{v}^{\rm eff} \vec{\rho} \right) = \vec{\mathcal{I}}[\rho] \tag{2}$$

with the effective velocity given by[2–4]

$$v_\lambda^{\rm eff} = \frac{(e')_\lambda^{\rm dr}}{(p')_\lambda^{\rm dr}}, \tag{3}$$

where $e(\lambda)$ denotes the bare energy. Using the Bethe equations, we have $\hat{v}^{\rm eff} \vec{\rho} = \frac{1}{2\pi} \hat{n} \vec{e'}^{\rm dr}$, so the derivatives read

$$\begin{aligned} \partial_t \vec{\rho} &= \frac{1}{2\pi} \left( \hat{1} + \hat{n}\hat{T}^{\rm dr} \right) \partial_t \hat{n} \vec{p'}^{\rm dr}, \\ \partial_x \left( \hat{v}^{\rm eff} \vec{\rho} \right) &= \frac{1}{2\pi} \left( \hat{1} + \hat{n}\hat{T}^{\rm dr} \right) \partial_x \hat{n} \vec{e'}^{\rm dr}, \end{aligned} \tag{4}$$

where we have defined the dressed kernel $\hat{T}^{\rm dr} = (\hat{1} - \hat{T}\hat{n})^{-1}\hat{T}$. Introducing the operator $\hat{R}$ given by $\hat{R} = \hat{1} - \hat{n}\hat{T}$, we have $\hat{1} + \hat{n}\hat{T}^{\rm dr} = \hat{R}^{-1}$, so that equation (2) becomes

$$\frac{1}{2\pi} \hat{R}^{-1} \left( \partial_t \hat{n} + \hat{v}^{\rm eff} \partial_x \hat{n} \right) \vec{p'}^{\rm dr} = \vec{\mathcal{I}}[\rho] \tag{5}$$

Using $\rho_\lambda^{\rm tot} = \frac{1}{2\pi} \vec{p'}^{\rm dr}|_\lambda$, we have

$$\partial_t n_\lambda + v_\lambda^{\rm eff} \partial_x n_\lambda = I_\lambda[n], \tag{6}$$

with the modified collision integral $I_\lambda[n] \equiv \frac{1}{\rho_\lambda^{\rm tot}} \left( \hat{R} \vec{\mathcal{I}}[\rho] \right) \Big|_\lambda$.

## II. NUMERICAL SOLUTION OF THE BOLTZMANN EQUATION

Motivated by the method of characteristics widely used in the context of PDEs and by Ref.[5] in the integrable case, we propose a solution of Eq (2) based on a finite element, backward (implicit) first order scheme solution given by $n_\lambda(x,t) = n_\lambda(x - v^{\text{eff}}[n(x,t)]\Delta t, t - \Delta t) + \Delta t I_\lambda[n(x,t)]$, which is obviously correct up to order $\mathcal{O}(\Delta t^2)$ and was found to be numerically stable in the integrable limit $I_\lambda = 0$ in Ref.[5]. A solution to this equation is found upon convergence when reducing the size of $\Delta t$, $\Delta x$ and $\Delta \lambda$. In practice we find quick convergence in most problems considered, and no further refinement of the method is needed. We remark however that going beyond first order in $\Delta t$ would not provide further insights into the phenomenology considered in this work, though it might be useful for future applications of our approach to access longer times. Solving the Boltzmann equation at time step $N_t$ requires an interpolation scheme to map out all possible arguments of the solution (characteristic curves) at time step $N_t - 1$, which we do by means of cubic splines.

In this implicit scheme, the right-hand side $n_\lambda(x - v^{\text{eff}}[n(x,t)]\Delta t, t - \Delta t) + \Delta t I_\lambda[n(x,t)]$ depends on the state at time $t$, and as such the whole equation must be solved iteratively, which in practice does not pose a problem reaching convergence quickly. Finding the effective velocity $v^{\text{eff}}$, as well as the collision term, which in principle is arbitrary, requires solving integral equations for the dressed quantities and the quasiparticle densities, only bounded by the discretization of $x$-space, the upper rapidity cutoff $\Lambda$, and the number of species of quasiparticle excitations. It is thus imperative to find an efficient way to carry out all (improper) integrals. Using a Legendre-Gauss quadrature scheme allows us to find convergence in all these for up to the times considered in this work using as little as $N_\lambda = 100$ points in rapidity space in the simplest instances with $\Lambda = 20$, and up $N_\lambda = 350$ in the most numerically demanding cases.

### A. Solution of the Boltzmann equation with GRTA

In principle the aforementioned scheme works generically for any collision integral. In this work we are mostly interested in the specific choice given by the GRTA, Eq (3) of the main text $\mathcal{I}_\lambda[\rho] = -(\rho_\lambda - \rho_\lambda^{\text{Gibbs}})/\tau$ (with the counterpart version in normal modes written as in (6)). As explained in the main text, $\rho_\lambda^{\text{Gibbs}}$ is the Gibbs state with the same value of the charge densities $q_\alpha(x,t)$, $\alpha = 1, ..., N$ as the state $\rho_\lambda$, so that $\frac{d}{dt}\int dx q_\alpha = 0$. As such, it is a non-linear functional of $\rho_\lambda$ and thus makes the right-hand side far from trivial. Its numerical implementation is straightforward: the first step is to discretize the Lagrange multipliers involved in the problem (e.g. in the case where the GRTA conserves particle number $Q_0 = N$ and energy $Q_2 = E$, these will be $\mu$ and $\beta$). For each set of Lagrange multipliers $\{\beta_\alpha\}$ we compute the associated conserved charge densities $\{q_\alpha\}$ via the equation of states known from TBA, $\{q_\alpha\} = f(\{\beta_\alpha\})$[6]. This yields a multidimensional grid for each $q_\alpha$. Since in all physical cases of interest $N \leq 3$, this first step can be solved very efficiently. In practice, it helps to adapt the range of this grid to the initial state under investigation. The second step is to compute at each time step and position $x$ the various charge densities of the conserved charges $q_\alpha = \int d\lambda \rho_\lambda h_\alpha(\lambda)$, and find the corresponding Lagrange multipliers that are closest to the ones used when computing the charge density grids. This can be further improved by interpolation of the grid, and we have also tried steepest-descent schemes. This gives us $\rho_\lambda^{\text{Gibbs}}$ at each space-time coordinate and thus the GRTA collision integral $I_\lambda$. We then follow the same steps indicated above to solve the GHD Boltzmann equation.

We have illustrated this approach in the main text for both interacting Bose gases and spin chains. In the case of the Lieb-Liniger model our simulations were obtained setting the coupling constant $c = 1$ and $m = 1/2$. For the far-from-equilibrium results we prepared the initial state to a $T = 2/3$ temperature background with a temperature excess of value $T = 20$ at the origin. We obtained converged results using $N_\lambda = 200$ and $\Delta t = 0.02$, $\Delta x = 0.75$. For the linear response regime we prepared the initial state with a local excess of temperature $T = 1.1$ on top of a background at temperature $T = 1$. We find convergence using $N_\lambda = 200$, $\Delta t = 0.05$ and $\Delta x = 0.5$. For the XXZ model (see below for more details) the parameters used were $N_\lambda = 300$, $\Delta t = 0.2$ and $\Delta x = 0.075$.

### B. Hydrodynamics of non-integrable Bose gases

To illustrate the GRTA approach, we study the crossover from generalized to conventional hydrodynamics in one-dimensional Bose gases, governed by the Lieb-Liniger Hamiltonian

$$\hat{H}_0 = \int dx\, \hat{\Psi}^\dagger \left(-\frac{\nabla^2}{2m} - \mu\right) \hat{\Psi} + c\hat{\Psi}^\dagger \hat{\Psi}^\dagger \hat{\Psi} \hat{\Psi}, \qquad (7)$$

with $m = 1/2$ and $c = 1$ hereafter. We first consider integrability-breaking perturbations that relax momentum: in this case, the conserved quantities in the Gibbs state of the GRTA are particle number and energy. We implement

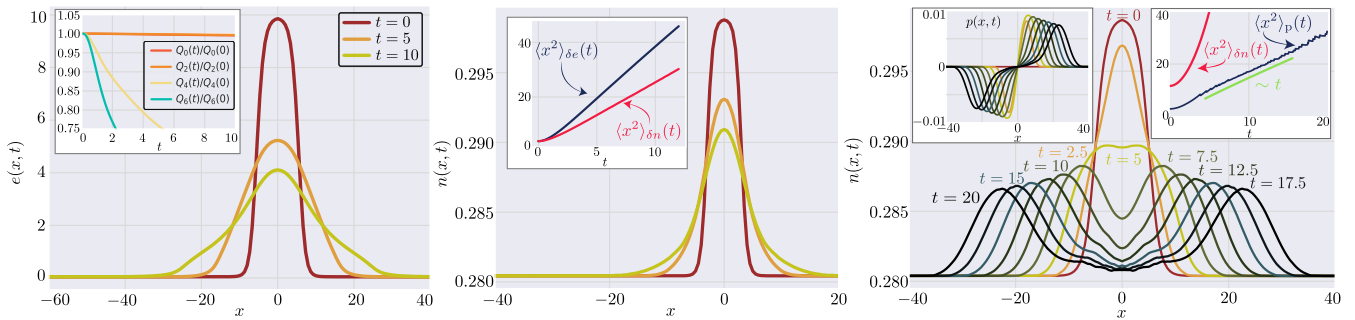

FIG. 1. **Crossover from generalized to conventional hydrodynamics** in 1d Bose gases using GRTA with $\tau = 1$. *Left*: Energy density *vs* time for free expansion of a cloud into vacuum, for a GRTA perturbation conserving energy and particle number. *Inset*: evolution of the charges $Q_n$ of the Lieb-Liniger model, showing conservation of particle number $Q_0 = N$ and energy $Q_2 = E$. *Middle*: Particle density after a linear-response perturbation to a thermal state with $T = 1$ and $\mu = 0$, for a GRTA perturbation conserving energy and particle number. *Inset*: variances of the energy and particle density profiles, showing diffusive behavior. *Right*: Linear-response initial state for a GRTA perturbation conserving energy, particle number and momentum. *Left inset*: Momentum profiles. *Right inset*: Variance of the particle number profiles showing ballistic transport (red), and diffusive broadening of the sound peaks in the momentum profiles (blue).

both far from equilibrium free expansions into vacuum of a cloud of atoms which models experiments on ultracold Bose gases (Fig. 1a)[7–18], and linear response setups where the initial state is a small local perturbation on top of an equilibrium Gibbs state (Fig. 1b). We confirm that the conservation of both energy and particle number are satisfied to a very good accuracy for all plotted time scales ($< 0.5\%$). We find that while the variance of the profiles of the local perturbation in both energy and particle density grow quadratically (indicating ballistic transport) in the integrable case, they crossover to linear (diffusive) growth for times $t \gg \tau$. Diffusive hydrodynamics is expected as momentum is not conserved, and we see that energy and particle number have different diffusion constants, inherited from the different Drude weights of the integrable limit.

We have also solved the hydrodynamic evolution of a Bose gas (7) with a perturbation that preserves particle number, energy, and momentum. Our scheme fully preserves Galilean invariance, so the particle current is momentum and is therefore conserved: we observe "sound modes" propagating ballistically in the nonintegrable case, which broaden diffusively on the time scales simulated. We also observe a small heat mode near the origin. This is consistent with what is expected from conventional, Navier-Stokes hydrodynamics in one dimension. We note that conventional hydrodynamics is generically anomalous in one dimension, and adding noise to our equations is expected to broaden the sound peaks in a superdiffusive way (dynamical exponent $z = 3/2$) – instead of diffusive – as predicted by the theory of nonlinear fluctuating hydrodynamics[19–21]. It would be interesting to include noise in our framework to check this.

## III. ENERGY TRANSPORT IN NON-INTEGRABLE QUANTUM SPIN CHAINS

We apply GRTA to the XXZ model with a staggered magnetic field along the transverse direction. We simulate the dynamics from first principles using the TEBD algorithm, as well as using the Boltzmann equation (6) with the GRTA approximation. All TEBD simulations are carried out by evolving the homogeneous infinite temperature density matrix in imaginary time with the inhomogeneous Hamiltonian

$$\hat{H} = \sum_i \frac{\beta(i + \frac{1}{2})}{\beta_0}(\hat{S}_i^x \hat{S}_{i+1}^x + \hat{S}_i^y \hat{S}_{i+1}^y + \Delta \hat{S}_i^z \hat{S}_{i+1}^z) + \frac{\beta(i)}{\beta_0} h_x (-1)^i \hat{S}_i^x, \tag{8}$$

and similarly for $\hat{V} = J'_x \sum_i (-1)^i \hat{S}_i^x \hat{S}_{i+1}^x$. We evolve using a 2nd order Trotter decomposition of the imaginary time evolution operator using Trotter steps of size $\Delta\beta = 0.001$ and number $\beta_0/\Delta\beta$, effectively attaining the far-from-equilibrium initial (inverse) temperature profile given by $\beta(x)$ in the homogenous Hamiltonian. We take the temperature profile to be a Gaussian centered on the middle bond of a length 200 chain:

$$\beta(x) = \beta_0 - (\beta_0 - \beta_M)e^{-(x-x_0)^2/L^2}, \tag{9}$$

with $\beta_0 = 0.5$, $\beta_M = 0.1$, and $L = 8$. We then time evolve the resulting initial state using a 4th order Trotter decomposition of the time evolution operator (of the homogeneous Hamiltonian) with Trotter step $\delta t = 0.4$. We set

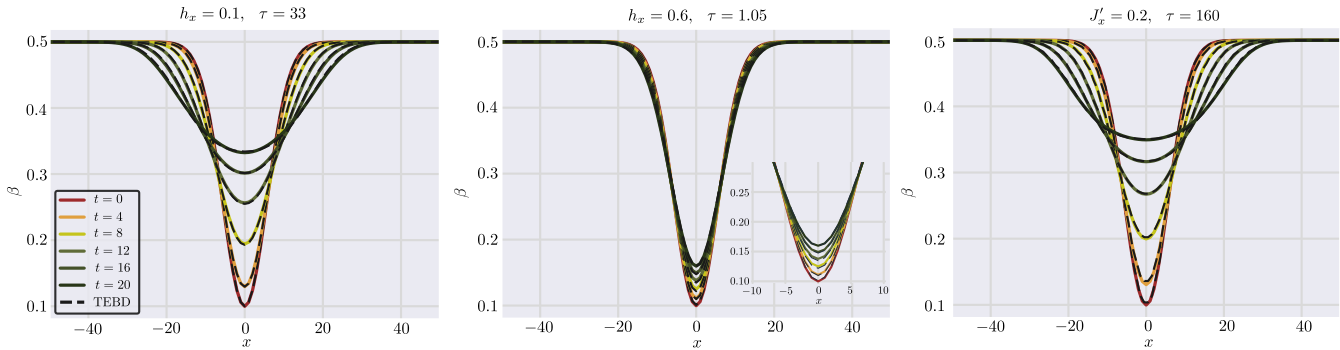

FIG. 2. **Integrability breaking for different perturbations in XXZ using TEBD and GRTA:** *Left:* Temperature profiles for the XXZ model with a staggered magnetic field $h_x = 0.1$, compared with that computed using GRTA for $\tau = 33$. *Middle:* Same as in left panel but with $h_x = 0.6$ and $\tau = 1.05$. *Right:* Temperature profile when the perturbation is given by a staggered $xx$−coupling with $J'_x = 0.2$ and compared with GRTA using $\tau = 160$.

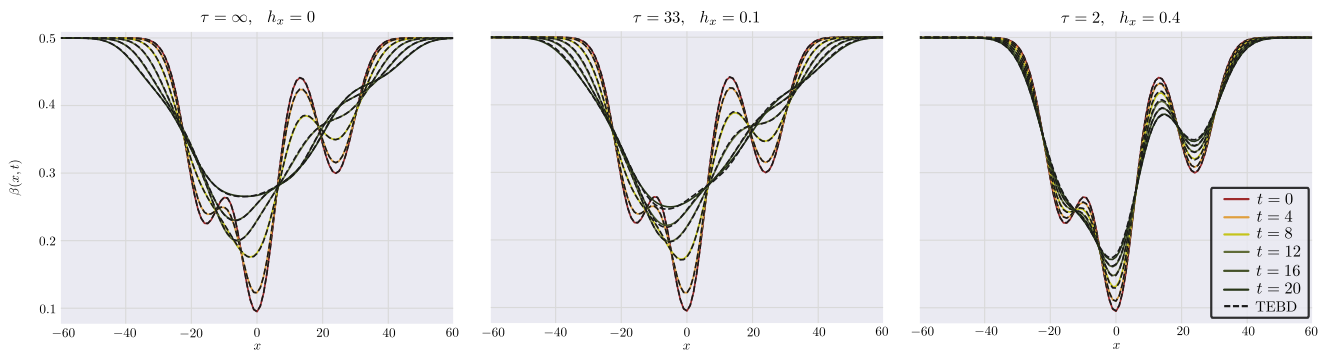

FIG. 3. **Generic energy transport in chaotic spin chains**: inverse temperature profiles $\beta(x,t) = 1/T(x,t)$ in an XXZ spin chain with a staggered transverse field $h_x$ breaking integrability, comparing TEBD and GRTA starting from a non-trivial inhomogeneous initial state. The values of $\tau$ in GRTA for each $h_x$ were determined from Fig. 2.

the maximum bond dimension $\chi_{\max} = 512$, and a truncation cutoff of $\epsilon = 10^{-9}$, which suffice to obtain converged results for all our simulations.

We find that GHD+GRTA treating $\tau$ as a free parameter agrees remarkably well with TEBD time evolution. The results for $h_x = 0.1$ and $h_x = 0.6$ (staggered $x$-fields) are shown in Fig. 2. Surprisingly, we find very good agreement even far away from the integrable point $h_x = 0$: for $h_x = 0.6$, $\tau \sim 1$ indicating a very quick crossover to diffusion. As shown in the main text, the fitted values of $\tau$ agree very well with the FGR scaling $\tau = C h_x^{-2}$, meaning that ultimately, non-equilibrium energy transport for any (reasonably small) value of $h_x$ can be inferred from fitting $C$ from a single instance of $h_x$. Time evolution predictions using those values of $\tau(h_x)$ for more involved initial states are shown in Fig. 3. Results for staggered $xx$-couplings are similar, as discussed in the main text (see also Fig. 2).



\end{document}